 \documentclass[final,5p,times,twocolumn]{elsarticle}

\usepackage{graphics}
\usepackage{graphicx}
\usepackage{epsfig}
\usepackage{color}

\usepackage{amssymb}
\usepackage{amsthm}

\biboptions{sort&compress}

\newcommand{\od}{\mathrm{d}}
\newcommand{\pd}{\partial}
\usepackage{enumerate,amsmath,bm,amssymb}

\journal{Chem. Phys. Lett.}

\begin{document}

\begin{frontmatter}



\title{
Numerical simulation method for Brownian particles dispersed in incompressible
fluids
}

\author[a1]{Hiroaki Yoshida\corref{cor1}}
\ead{h-yoshida@mosk.tytlabs.co.jp}
\author[a2]{Tomoyuki Kinjo}
\author[a2,a3,fn1]{Hitoshi Washizu}

\address[a1]{Toyota Central R\&D Labs., Inc., Bunkyo-ku, Tokyo 112-0004, Japan}
\address[a2]{Toyota Central R\&D Labs., Inc., Nagakute, Aichi 480-1192, Japan}
\address[a3]{Elements Strategy Initiative for Catalysts and Batteries (ESICB), Kyoto University, Kyoto 615-8245, Japan}
\cortext[cor1]{Corresponding author.}
\fntext[fn1]{Present address: Graduate School of Simulation Studies, University of Hyogo, 7-1-28 Minatojima-minamimachi, Chuo-ku, Kobe, Hyogo 650-0047, Japan.}





\begin{abstract}
We present a numerical scheme for simulating the dynamics of Brownian particles suspended in a fluid. The motion of the particles is tracked by the Langevin equation, whereas the host fluid flow is analyzed by using the lattice Boltzmann method. The friction force between a particle and the fluid is evaluated correctly based on the velocity difference at the position of the particle. The coupling method accurately reproduces the long-time tail observed in the velocity auto-correlation function. We also show that the fluctuation-dissipation relation holds between the relaxation of a single particle and the velocity autocorrelation function of fluctuating particles.
\end{abstract}

\begin{keyword}
Colloidal suspensions \sep
Brownian particles \sep
Langevin dynamics \sep
Lattice Boltzmann method


\end{keyword}

\end{frontmatter}



\section{\label{sec_intro}Introduction}
Colloidal suspensions of sub-micro or nano particles
play important roles in many situations.
Examples include processes of producing secondary-battery electrodes
and coats of paint~\cite{MFB+2019,AJS+2018}.
Recently, functional fluids that
change their rheological properties have also attracted attentions~\cite{KAM+2018,AMP+2018}.
The particles suspended in solutions 
in those systems thermally fluctuate,
which is referred to as the Brownian motion.
In order to understand the connection
between this microscopic motion of the particles
and the macroscopic fluid properties, 
the hydrodynamic motion of the solution
induced by the Brownian particles
has to be comprehended correctly.

The computational method
that is most widely used for simulating
the Brownian motion of particles is the one
solely based on the Langevin equation,
referred to as the Brownian dynamics simulation.
The simplest approach
in implementation of the Brownian dynamics simulation
assumes that the solvent
is at rest and not affected by the motion of the particles,
where random force representing the thermal fluctuation
and an effective friction force in proportion to the particle velocity
exert on a particle. 
To include the hydrodynamic interaction effect into the fluctuation and friction forces
is still challenging despite a number of attempts that have been made to incorporate the effect of the motion of the surrounding fluid~\cite{EM1978,BB1988,NL2009,SCT2011,Maxey2017,RR2018,PPK+2018,LZC+2019,LYH2019},
because simulating directly the motion of the fluid is required to capture the hydrodynamic interaction effect
that stays within the fluid at the time scale of 
the momentum dissipation.
There exist
two types of simulation methods
in which the fluid flow 
is directly simulated along with
tracking the motion of the particles:
(i) a particle has finite size
comparable with the computational domain for the fluid flow,
and a boundary condition at the surface of the 
particle is imposed in the flow simulation~\cite{NY2005,IY2009,LV2001,OKCO2008,Sman2010,MLH2017},
(ii) a particle is represented by a single point, and a model 
friction force is employed in order to incorporate the interaction
between the particle and the fluid~\cite{AD1998,AD1999,LD2004,CH2005}.
The latter is advantageous from the point of view of computational cost,
but the inaccuracy of the local estimation of the friction force can be
a problem in certain physical situations.

In this work,
we present an accurate local estimation of the fluid-particle friction force.
Specifically, the friction force acting on a particle is estimated by fitting the analytical solution 
for the flow around a Stokes-let to the flow field obtained numerically.
The reaction force acts on the position of the particle,
which realizes two-way coupling between the particle and fluid motions.
The lattice Boltzmann method is employed for the flow simulation,
which is compatible with massive parallel computing,
and is easy to apply various types of boundary conditions
such as the periodical shear boundary,
and complex structure of obstacles.
For validation of the method, the long-time tail observed 
in the velocity auto-correlation function (VACF) is compared with the
analytical expression. The fluctuation--dissipation theorem,
which relates the VACF and the relaxation process of the velocity and 
acceleration of a single particle, is also examined.

%
\section{\label{sec_method}Numerical algorithm}

\subsection{\label{sec_coup}Estimation of the friction force}

We first describe the equation governing the motion of Brownian
particles, and present the algorithm for coupling
with the fluid flow.
The motion of the particles is described by the Langevin equation
in the following form:
\begin{align}
&M_i\dfrac{\od\bm{V}_i}{\od t}=\bm{F}_{Ci}+\bm{F}_{Hi}+\bm{F}_{Ri},
\label{s2-langevin}\\
&\bm{V}_i=\dfrac{\od \bm{r}_i}{\od t},
\label{s2-langevin2}
\end{align}
where $M$ is the mass, $\bm{r}_i$ is the position, and $\bm{V}_i$ is the velocity,
of $i$th particle.
In Eq.~\eqref{s2-langevin},
the conservative inter-particle force, 
the dissipation force, and the fluctuation force
are denoted by $\bm{F}_{C}$, $\bm{F}_{H}$, and $\bm{F}_{R}$, respectively.
{\color{black}
In the present study, 
we employ the conservative force
derived from the potential
of conventional dissipative particle dynamics~\cite{GW1997}
having the following form:
}
\begin{equation}
U(\bar{r}_{ij})=a_f\left(\bar{r}_{ij}-\dfrac{1}{2}\bar{r}_{ij}^2\right),
\label{s2-softpot}
\end{equation}
where $a_f$ is a coefficient determining the intensity of the
inter-particle force, and $\bar{r}_{ij}$ is the inter-particle distance.
If the particle density is not significantly large, 
the choice of $\bm{F}_C$ has limited effect on the results,
because the hydrodynamic interaction through the fluid flow acts as a
repulsive force.
{\color{black}
The fluctuation force meets the following property:
}
\begin{align}
&\langle F_{Ri}^p(t,\bm{r}_i)\rangle=0,
\label{s2-rf1}\\
&\langle F_{Ri}^p(t,\bm{r}_i)F_{Ri}^q(t',\bm{r}'_i)\rangle=\sigma^2\delta(p-q)\delta(t-t')\delta(\bm{r}_i-\bm{r}'_i),
\label{s2-rf2}
\end{align}
where $\langle\cdot\rangle$ denotes the ensemble average, $\delta$
is the Kronecker delta, and $\sigma$ is a constant determining the
intensity of the random force. Here, superscripts $p$ and $q$ indicate the
Cartesian components of the force.

In the simplest Brownian dynamics simulation,
the dissipation force is proportional to the velocity of the particle
itself $\bm{V}(t)$:
\begin{equation}
\bm{F}_{H}=-\gamma\bm{V}(t),
\label{s2-dis0}
\end{equation}
where $\gamma$ is the constant for the friction force
related to the fluid viscosity via Stokes' law: $\gamma=6\pi\eta R$
with $R$ representing the particle radius.
Since Eq.~\eqref{s2-dis0} assumes the fluid to be at rest,
the momentum transport through the fluid is neglected,
and thus the Brownian motion of particles 
at the time scale of the momentum dissipation is not
described accurately.
For example, 
the VACF $C_{v}(t)=\langle \bm{V}(0)\cdot\bm{V}(t)\rangle$,
which should exhibit the so-called long-time tail
described by a power function of time ($C_{v}(t)\propto t^{-3/2}$),
decays exponentially if we use Eq.~\eqref{s2-dis0} ($C_{v}(t)\propto \exp(-\gamma t/M)$),
as will be discussed later in Sec.~\ref{sec_numerical}.
In order to capture the hydrodynamic interaction effect,
D\"unweg and his co-workers proposed 
the friction force to be in proportion to the
particle velocity relative to the fluid velocity at the position of the
particle, which is obtained from an independent simulation of the
Navier--Stokes equations:~\cite{AD1998,AD1999}
\begin{equation}
\bm{F}_{Hi}=-\gamma(\bm{V}_i(t)-\bm{u}(t,\bm{r}_i)),
\label{s2-dis1}
\end{equation}
where $\bm{u}$ is the velocity field of the fluid.
This correction certainly realizes a momentum transport through the
fluid, and the long-time tail is qualitatively reproduced.
However, the underestimation of the friction force due to
the local estimation results in an insufficient accuracy
in reproducing the VACF, as they recognized, and the fluctuation-dissipation theorem is 
thus not satisfied without an empirical tuning of the coefficient $\gamma$.

In the present study, we replace the flow velocity $\bm{u}$ in
Eq.~\eqref{s2-dis1}, i.e., the flow velocity of the fluid at the
particle position, by the flow velocity away from the position of the particle
$\bm{u}^{\infty}$, still estimated locally.
Since the flow velocity away from the particle
is not unique, we use the value of $\bm{u}^{\infty}$
obtained assuming that the flow velocity field around the particle is
approximated by the one-way flow field around a spherical particle.
Since, in the actual numerical simulation,
the flow velocity is obtained
at a regular computational grid by using the lattice Boltzmann method,
the value of $\bm{u}^{\infty}$ is evaluated using 
the flow velocity at the neighboring grid points
$\bm{u}^{\mathrm{N}}_p=\bm{u}(\bm{x}_p)$,
where $\bm{x}_p$ ($p=1,\ldots,8$) represents the neighboring eight grid points around the particle.

To this end, we use the following
analytical solution of the flow past a Stokes-let with the intensity $6\pi\eta R U^{\infty}$:
\begin{align}
&\bm{u}^{\mathrm{A}}=U^{\infty}\left[\left(1-\dfrac{3}{4}\dfrac{R}{r}\right)\bm{e}_x
-\dfrac{3Rx}{4r^3}\bm{x}\right].
\label{s2-slet-v}
\end{align}
Here, the flow in the $x$-direction is assumed, and the Stokes-let is at
the origin ($\bm{e}_x$ is a unit vector in the $x$-direction, and $\bm{x}$ is the spatial coordinate,).
After transforming Eq.~\eqref{s2-slet-v}
such that the origin is at the particle position, and the flow 
is in the direction of the vector $\bm{u}^{\mathrm{N}}_p-\bm{V}$
averaged over eight points (unit vector in this direction is denoted by $\bm{e}^{\mathrm{N}}$),
we calculate the difference between the relative flow velocity
$\bm{u}^{\mathrm{N}}_p-\bm{V}$ at grid points and the $\bm{u}^{\mathrm{A}}$
(Eq.~\eqref{s2-slet-v}) at corresponding points.
{\color{black}
The value of $U^{\infty}$ is then determined to minimize the
sum of the squares of the difference, 
$\sum_{p=1}^8|\bm{u}^{\mathrm{N}}_p-\bm{V}-\bm{u}^{\mathrm{A}}(\bm{x}_p)|^2$.
Using the value of $U^{\infty}$ determined, the flow velocity away from the particle is obtained as $\bm{u}^{\infty}=U^{\infty}\bm{e}^{\mathrm{N}}+\bm{V}$.
}

In the flow simulation by means of the lattice Boltzmann method,
the momentum due to the motion of the particle is transferred 
via the reaction force of $\bm{F}_H$:
\begin{equation}
\bm{F}_{pi}=-\bm{F}_{Hi}\delta(\bm{x}-\bm{r}_i)/\Delta V,
\label{s2-force}
\end{equation}
where $\Delta V$ is the unit cell volume of the computational grid system.
The pointwise force is distributed around the neighboring grid points 
with a weight depending on the distance from $\bm{r}_i$.

\subsection{\label{sec_lbm}Outline of the lattice Boltzmann method}

In the present study, we employ the
lattice Boltzmann method (LBM) to
obtain the flow velocity $\bm{v}(t,\bm{x})$ and the pressure $p(t,\bm{x})$
governed by the Navier--Stokes equations:
\begin{align}
&\nabla\cdot \bm{v}=0,
\label{s2-divfree}\\
&\dfrac{\pd \bm{v}}{\pd t}+(\bm{v}\cdot\nabla)\bm{v}=
-\dfrac{1}{\rho_0}\nabla p + \nu \nabla^2\bm{v} + \dfrac{1}{\rho_0}\sum_i\bm{F}_{pi},
\label{s2-ns}
\end{align}
where $\rho_0$ is the reference fluid density and $\nu$ is the kinetic viscosity.

In the LBM,
the flow behavior is described in terms of the 
velocity distribution function $f_{\alpha}(t,\bm{x})$, 
instead of directly handling the variables $\bm{v}$ and $p$.
Here, $\alpha=0,1,2,\ldots,n$ with $n$ being the
number of discrete velocities.
The value of $f_{\alpha}$ represents the partial
fluid density,
each of which travels over the regular lattice
with the discrete velocity assigned.
The direction of the discrete velocity is defined in terms of the
vector $\bm{e}_{\alpha}$.
There are several sets of the discrete velocities, which 
satisfy the constraints that must be satisfied in order to reproduce the
Navier--Stokes equations.
We here employ the fifteen-velocity set, one
of the most widely used sets (see, e.g., Ref.~\cite{S2001} for the specific expression).
The relation between the local density of the fluid and the velocity
distribution function is:
\begin{equation}
\tilde{\rho}=\sum_{\alpha}f_{\alpha},
\label{s2-density}
\end{equation}
where $\tilde{\rho}$ is the density normalized by $\rho_0$.
Since the incompressible fluid is considered,
$\tilde{\rho}$ should be close to unity, and thus the value of
$f_{\alpha}$ is regarded as a discrete probability distribution function.
Therefore, the velocity of the fluid is expressed as:
\begin{equation}
\bm{u}=\sum_{\alpha}C\bm{e}_{\alpha}f_{\alpha},
\label{s2-fv}
\end{equation}
where $C$ is the speed defined in terms of the time step $\Delta t$ and 
the grid interval $\Delta x$ as $C=\Delta x/\Delta t$.
{\color{black} 
The basic equation of the LBM then reads
\begin{align}
&f_{\alpha}(t+\Delta t,\bm{x}+\bm{e}_{\alpha}\Delta
 x)=f_{\alpha}(t,\bm{x})
+ Q_{\alpha}[f](t,\bm{x}) +\Delta t \omega_{\alpha}G_{\alpha},
\label{s2-lbe}
\end{align}
where $Q_{\alpha}$ is the collision operator that defines interaction between $f_{\alpha}$'s:
\begin{equation}
Q_{\alpha}[f]=\dfrac{1}{\tau}\left[f^{\mathrm{eq}}_{\alpha}(\tilde{\rho}(t,\bm{x}),\bm{u}(t,\bm{x}))-f_{\alpha}(t,\bm{x})\right].
\label{s2-collision}
\end{equation}
}Here, $\tau$ is the relaxation-time coefficient in relation with
the fluid viscosity:
\begin{align}
\nu=\dfrac{1}{3}\left(\tau-\dfrac{1}{2}\right)\dfrac{\Delta x^2}{\Delta t},
\label{s2-ident3}
\end{align}
and $f^{\mathrm{eq}}_{\alpha}$ is the equilibrium distribution function
defined as
\begin{align}
&f^{\mathrm{eq}}_{\alpha}(\tilde{\rho},\bm{u})=\omega_{\alpha}\bigg[\tilde{\rho}+\dfrac{3}{C}u_je_{j\alpha}
+\dfrac{9}{2C^2}(u_j e_{j\alpha})^2-\dfrac{3}{2C^2}{u_j}^2\bigg],
\label{s2-eq}
\end{align}
where $\omega_{\alpha}$ is a weight coefficient, of which the specific expression
is dependent on the set of discrete velocities (cf. Table~5.1 of Ref.~\cite{S2001}).
The interaction force $\bm{F}_{pi}$ acts on the fluid through
$G_{\alpha}$ defined as
\begin{equation}
G_{\alpha}=\dfrac{3F_{pj}e_{\alpha j}}{\rho_0C}.
\label{s2-force}
\end{equation}

We summarize the computational process of the LBM.
Given the distribution function at $t$,
$f_{\alpha}(t+\Delta t,\bm{x})$ is obtained 
through two steps, namely,\\
\noindent (i) Collision process:
\begin{equation}
\hat{f}_{\alpha}(t,\bm{x})=
f_{\alpha}(t,\bm{x})+Q_{\alpha}[f](t,\bm{x})
+\Delta t \omega_{\alpha}G_{\alpha}.
\label{s2-coll}
\end{equation}
\noindent (ii) Streaming process:
\begin{equation}
f_{\alpha}(t+\Delta t,\bm{x}+\bm{e}_{\alpha}\Delta x)=\hat{f}_{\alpha}(t,\bm{x}).
\label{s2-trans}
\end{equation}
Then, we calculate the physical quantities $\tilde{\rho}$ and $\bm{u}$ 
using Eqs.~\eqref{s2-density} and \eqref{s2-fv}.
The pressure $p$ is expressed in terms of the local density
$\tilde{\rho}$ as $p={C^2}\rho_0\tilde{\rho}/3$.
It is proven that the above process yields an approximated solution to
the incompressible Navier--Stokes equation~\cite{JKL2005,JY2005,YH2014}.

{\color{black}
To conclude this section, we remark a few points on the parameter ranges and extensions.
Firstly, since the relative velocity $U^{\infty}$ between the flow and a particle 
is computed using the neighboring grid points around the particle,
it should not cover all these neighboring grid points
i.e., we should ensure that $R \lesssim \sqrt{3}\Delta x$. Otherwise
the possible modification is to use the other grid points than the nearest points in estimation of $U^{\infty}$,
say eight corners of $(2\Delta x)^3$ cube. 
The second remark is on the density of the particle suspension.
For the same reason for the estimation of $U^{\infty}$, the spacing between particles should not always be small,
requiring the density not being too large; we have checked the present method works correctly for the volume fraction below $0.1$.
Finally we remark on the possibility of including particle rotations.
Since the flow field around a rotating sphere decays more quickly than the one considered herein,
the approximation of the present method should be sufficient in most cases to capture the basic properties of suspensions.
Nevertheless, this extension would be possible with estimating the rotational friction using a Stokes flow solution around a rotating sphere,
with the similar technique for Eq.~\eqref{s2-slet-v}.
}

\begin{figure}[t]
\begin{center}
\vspace*{0mm}
\includegraphics[scale=0.65]{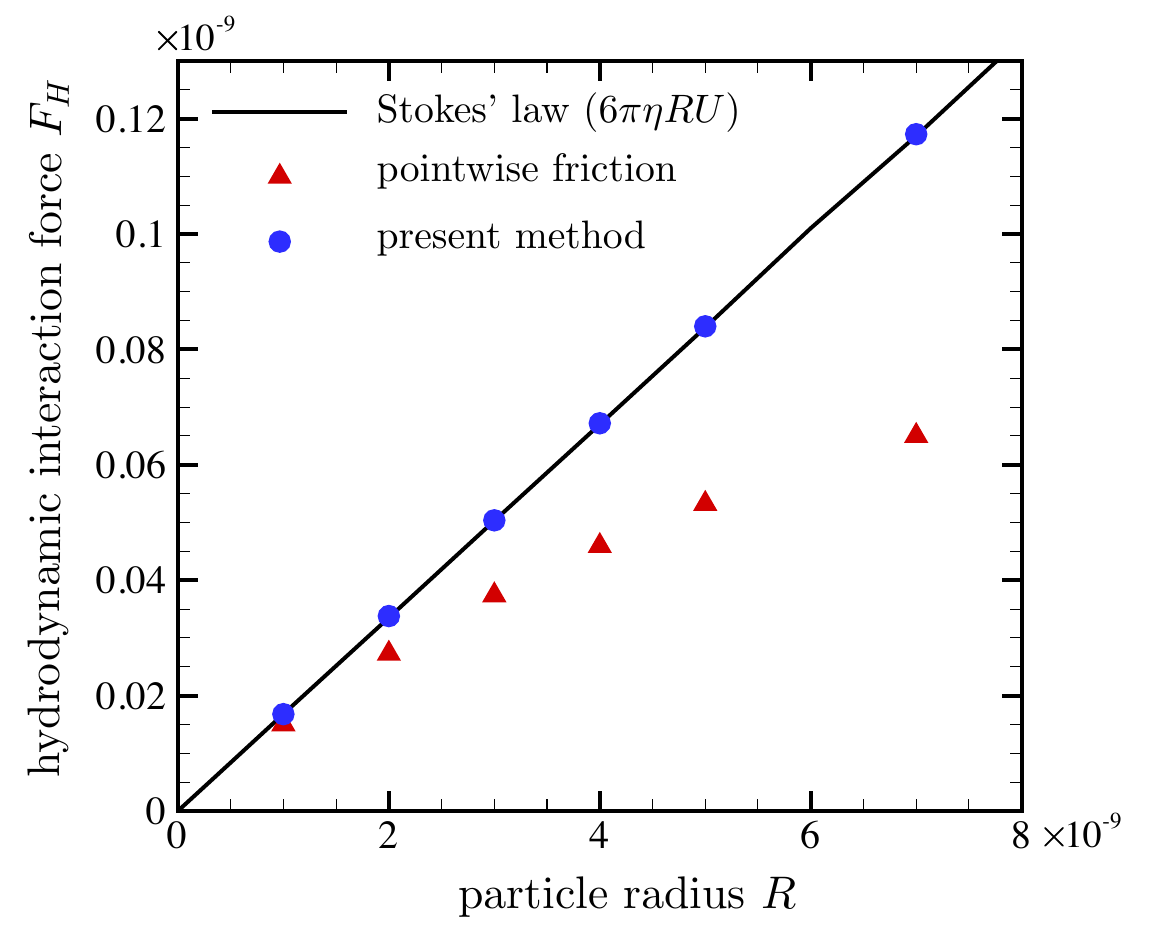}
\vspace*{-5mm}
\caption{Hydrodynamic interaction force $F_H$ acting on a sphere at rest in a uniform flow $U$.
The symbol {$\blacktriangle$} indicates the results obtained using the pointwise friction model \eqref{s2-dis1},
whereas the symbol {\Large $\bullet$} indicates the results of the present method.
The solid line is Stokes's formula $F_H=6\pi\eta R U$, where $\eta=\rho_0\nu$. 
$U=1$ [m/s], $\nu=0.89\times 10^{-6}$ [m$^2$/s], and $\rho_0=10^3$ [kg/m$^3$].
The size of the simulation box is $500\times 10^{-9}$ [m], 
the grid interval for the lattice Boltzmann method is $\Delta x=5\times 10^{-9}$,
and the time step is $\Delta t=0.5\times 10 ^{-12}$ [s].}
\label{fig01}
\end{center}
\end{figure}  
%

%
%
\section{\label{sec_numerical}Numerical results}

The existing model given in
Eq.~\eqref{s2-dis1} fails to accurately reproduce 
the friction force acting on a spherical particle
on which the stick boundary condition is imposed, as mentioned in
Sec.~\ref{sec_coup}.
In order to confirm the
improvement of the present method,
we here consider a very simple problem
of a flow past a single spherical particle
that is fixed ($\bm{V}(t)=0$).
Figure~\ref{fig01} plots
the force acting on the particle
as a function of the particle radius.
When the Reynolds number is defined as $\mathrm{Re}=UR/\nu$,
with $U$ being the given flow velocity at infinity,
is small, the friction force is a linear function
of the radius, $F_H=6\pi\eta R U$ (Stokes' law).
Since the relative velocity $|\bm{u}(t,\bm{r})|$ estimated at the position of
the particle is obviously smaller than $U$,
the friction is underestimated by Eq.~\eqref{s2-dis1}.
On the contrary, the present method estimates locally,
with the aid of the flow velocity $\bm{u}^{\mathrm{N}}_p$ near the particle,
a value of $\bm{u}^{\infty}$ close to $U$, 
hence the model evaluates correctly the friction force
which agrees well with Stokes' law.

\begin{figure}[t]
\begin{center}
\vspace*{0mm}
\includegraphics[scale=0.65]{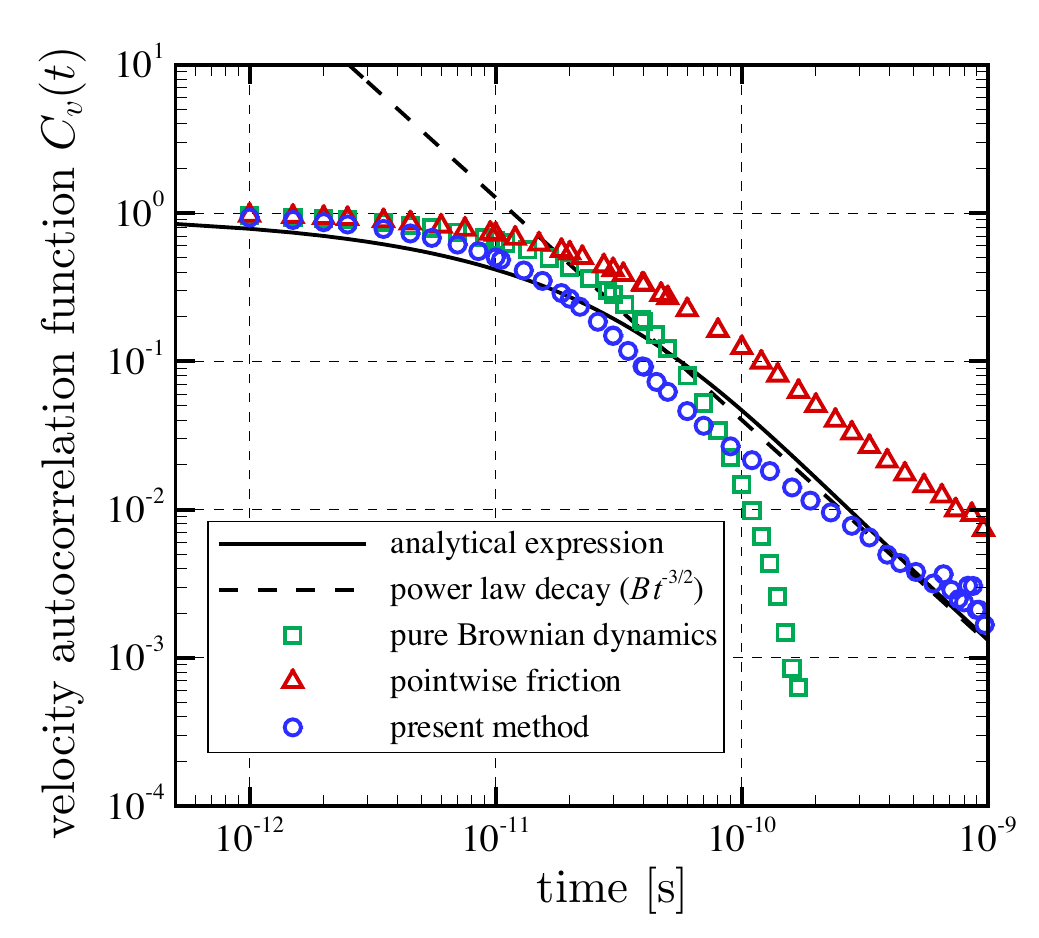}
\vspace*{-3mm}
\caption{Velocity autocorrelation function $C_v(t)$ normalized by $C_v(0)$ for fluctuating Brownian particles.
The suspension consists of $500$ identical particles with $R=5\times 10^{-9}$ [m] and $M=2\times 10^{-21}$ [kg].
Other conditions are the same as Fig.~\ref{fig01}.
The symbol {$\square$} indicates the result of the pure Brownian dynamics (Eq.~\eqref{s2-dis0}),
the symbol {$\vartriangle$} indicates the result obtained using the pointwise friction model (Eq.~\eqref{s2-dis1}),
and the symbol {\Large $\circ$} indicates the result of the present method.
The solid line is the analytical expression for $C_v(t)$ \cite{Hinch1975,PP1981,PL2006},
and the dashed line is its long-time limit ($C_v(t)/C_v(0)\sim Bt^{-3/2}$
 with $B=M/[12\rho_f(\pi \nu)^{3/2}]$).}
\label{fig02}
\end{center}
%

%
\begin{center}
\includegraphics[scale=0.65]{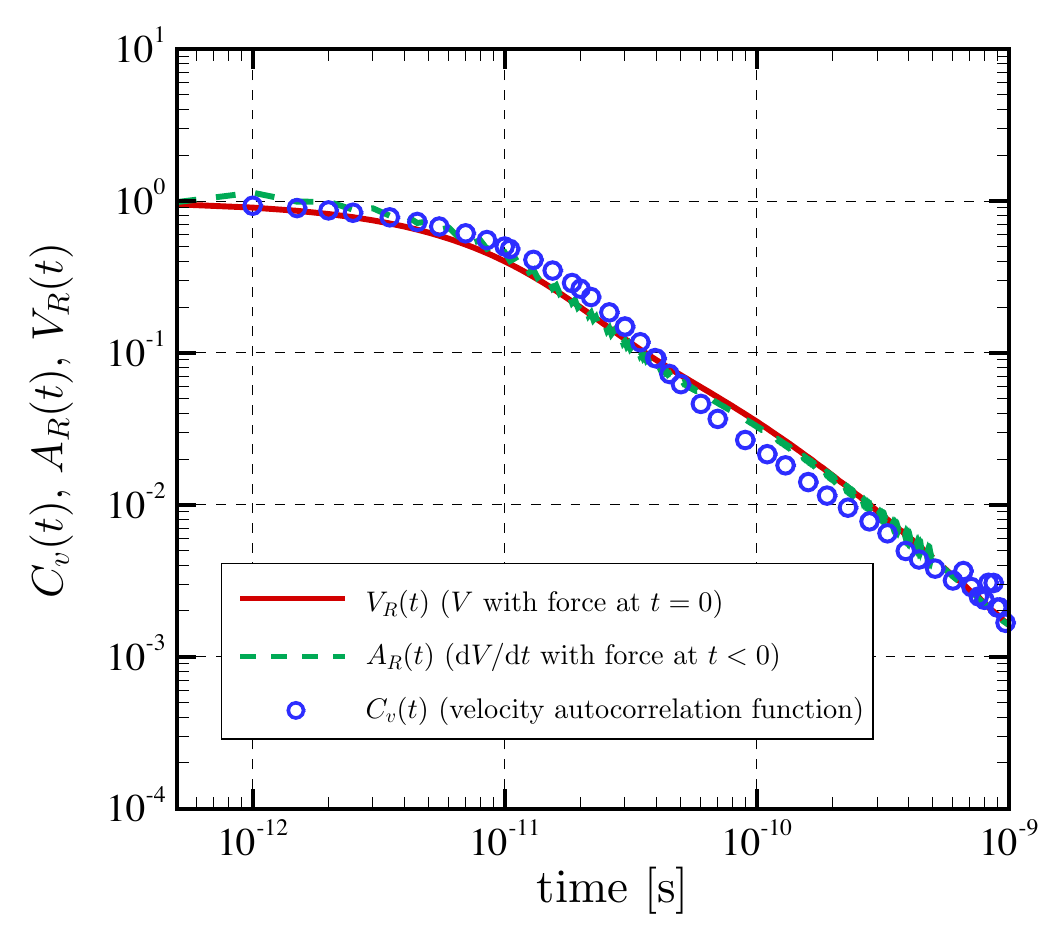}
\vspace*{-3mm}
\caption{Velocity autocorrelation function $C_v(t)$ ({\Large $\circ$}) compared to
response function $V_{R}(t)$ (solid line) and relaxation function $A_{R}(t)$ (dashed line).}
\label{fig03}
\end{center}
\end{figure}  

We next consider a system with
dispersed particles in thermal equilibrium,
to examine the VACF $C_v(t)$ of the particles.
Whereas the usual Brownian dynamics with the friction force given by Eq.~\eqref{s2-dis0}
yields a $C_v(t)$ decaying exponentially with respect to $t$,
the memory effect due to the interaction with the surrounding fluid
makes $C_v(t)$ decay more slowly, showing the long-time tail.
The generalized Langevin equation~\cite{Kubo1966} supplemented with the
friction force at a small Reynolds number gives
an analytical expression of $C_v(t)$, which 
shows $C_v(t)/C_v(0)\sim Bt^{-3/2}$ as $t\to\infty$, where $B=M/[12\rho_f(\pi \nu)^{3/2}]$~\cite{Hinch1975,PP1981,PL2006}.
In Fig.~\ref{fig02},
along with the analytical expression, 
we plot the VACF $C_v(t)$ obtained using Eq.~\eqref{s2-dis0},
Eq.~\eqref{s2-dis1}, and the present method using $\bm{u}^{\infty}$.
Although the result of Eq.~\eqref{s2-dis1} reproduces the long-time tail,
the value of $C_v(t)$ still differs form the analytical expression.
On the other hand, 
the VACF obtained using the present method
exhibits better quantitative agreement with the analytical expression.
There is a slight discrepancy of the present method in the region around $t=10^{-10}$\,[s].
This is a numerical error inherent to the LBM due to 
a small pressure wave which originates from the artificial compressibility.

As a consequence of the fluctuation-dissipation theorem,
the relaxation processes of the velocity
and acceleration coincide with the VACF.
More precisely,
if we denote with $V_R(t)$
the history of the velocity $V(t)$ of a particle
kicked by an instant force at $t=0$,
and we denote with $A_R(t)$ 
the acceleration $\od V/\od t(t)$ of a particle
after a force exerted on the particle in $-\infty<t<0$
is released.
Then the relation $C_V(t)=V_R(t)=-A_R(t)$ holds
(both $V_R$ and $A_R$ are normalized by the values at $t=0$).
The relation is known as Onsager's regression hypothesis 
that was derived from the fluctuation-dissipation theorem
by Callen and Welton~\cite{CW1951}.
In Fig.~\ref{fig03}, we plot 
$V_R(t)$, $A_R(t)$ and $C_v(t)$ obtained using the present method
for the hydrodynamic interaction.
Fairly good agreement among the three functions confirms
that the thermal equilibrium state 
of particles in a solvent is correctly simulated
using the present coupling method.

\section{\label{sec_summary}Conclusion}
In the present study,
a numerical algorithm
for simulating the behavior of a fluid
with Brownian particles is presented,
in which 
the motion of the particles
is tracked by means of the Langevin equation,
while the fluid flow obeying the Navier--Stokes equations
is simulated using the LBM.
The relative velocity between the particle and the fluid
is evaluated from the local information, which improves 
the accuracy  of the hydrodynamic interaction force.
Several numerical simulations are performed to check the following features:
(1) the friction force acting on a pinned particle in a one-way flow
satisfies Stokes' law.
(2) the long-time behavior of the VACF predicted theoretically 
is reproduced correctly.
(3) the relaxation process under given disturbances agrees with
the VACF (Onsager's regression hypothesis).
The proposed algorithm yields correct hydrodynamic interaction force
when the particles are not close, or the number density is not very
large. The possibility of extending the present method
to highly dense suspensions is now under investigation.

\section*{Acknowledgments}
The authors are grateful to S. Iwai for computer assistance in preparing the manuscript.
Part of the work was supported by MEXT program ``Elements
Strategy Initiative to Form Core Research Center'' (since 2012).  (MEXT
stands for Ministry of Education, Culture, Sports, Science, and Technology, Japan.)






\end{document}